\documentclass[aps,prl,twocolumn,superscriptaddress,showpacs]{revtex4}

\usepackage{amsmath}
\usepackage{amsfonts}
\usepackage{amssymb}
\usepackage{graphics}
\usepackage{graphicx}

\begin{document}


\title{Is orbital angular momentum always conserved in spontaneous parametric down-conversion?}

\author{Sheng Feng} 
\email{sfeng@ece.northwestern.edu}
\affiliation{Center for Photonic Communication and Computing, EECS 
Department, Northwestern University, Evanston, IL 60208-3118, U.S.A.}
\author{Chao-Hsiang Chen}
\affiliation{Department of Physics and Astronomy, Northwestern 
University, Evanston, IL 60208-3112, U.S.A.}
\author{Geraldo A. Barbosa}
\affiliation{Center for Photonic Communication and Computing, EECS 
Department, Northwestern University, Evanston, IL 60208-3118, U.S.A.}
\author {Prem Kumar}
\affiliation{Center for Photonic Communication and Computing, EECS 
Department, Northwestern University, Evanston, IL 60208-3118, U.S.A.}
\affiliation{Department of Physics and Astronomy, Northwestern 
University, Evanston, IL 60208-3112, U.S.A.}




\date{\today}

\begin{abstract}
In the non-linear optical process of type-II spontaneous parametric down-conversion, we present on an experiment showing that the two-photon detection amplitude of the down-converted beams does not generally reproduce the transverse profile of the pump beam that carries non-zero orbital angular momentum. We explain this observation by that orbital angular momentum is not conserved in the type-II non-linear process due to the broken rotational symmetry of the Hamiltonian.
\end{abstract}


\maketitle



The physical variable of orbital angular momentum (OAM) can be used to prepare multi-dimensional entanglement \cite{mair01} and hyper-entanglement \cite{barreiro05}, which necessitate OAM conservation. Largely because of this, much attention has been attracted by the topic of OAM conservation in the process of spontaneous parametric down-conversion (SPDC) \cite{arlt99,arnaut00,arnold02,terriza03,walborn04}, where photon pairs entangled in OAM can be created. Similar studies were carried out in the case of stimulated down-conversion as well \cite{caetano02,huguenin06}. Nowadays, OAM conservation in SPDC process is widely assumed by researchers in this field \cite{vaziri02,molina04,ren04,oemrawsingh05,calvo07}. However, the fundamental ground of the origin of OAM conservation in SPDC process relating to the rotational symmetry of Hamiltonian has been very rarely touched. Given that type-II SPDC process lacks necessary rotational symmetry for OAM to be perfectly conserved, which will be explained soon, there is a chance to experimentally observe this symmetry-breaking-determined OAM non-conservation in this non-linear process. In this Letter, we present on an experiment, which was stimulated by the theoretical work of Arnaut and Barbosa \cite{arnaut00}, with novel observation, and propose an explanation that OAM is not conserved in the type-II SPDC process due to the broken symmetry of the Hamiltonian.

\begin{figure}[h]
 \begin{center}
 \unitlength = 1mm
 \begin{minipage}[c]{50mm}
  \begin{picture}(50,50)
    \hspace{-0.6in}
    \includegraphics[scale=0.10]{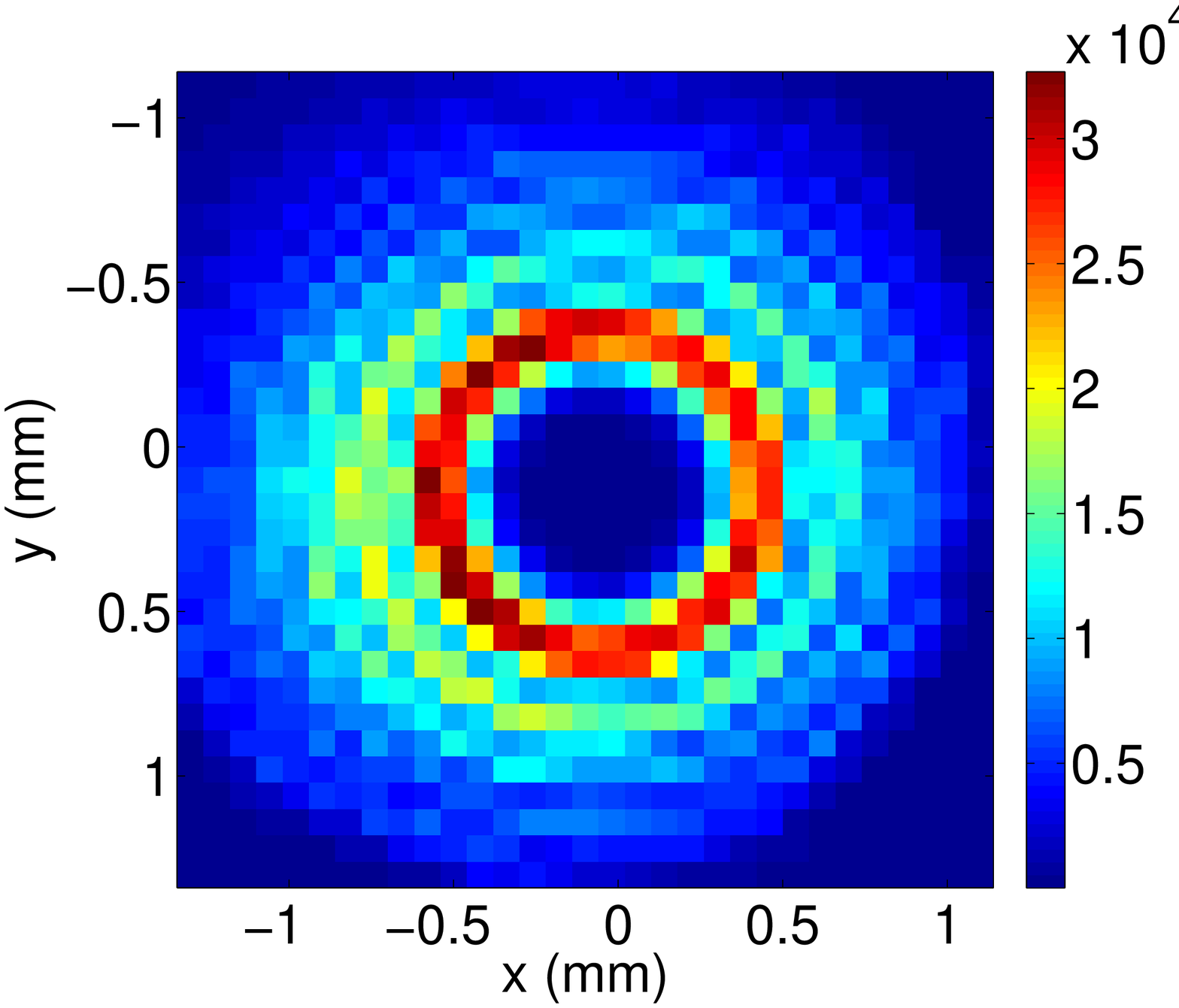}
    \hspace{-0.8in}
    \includegraphics[scale=0.4]{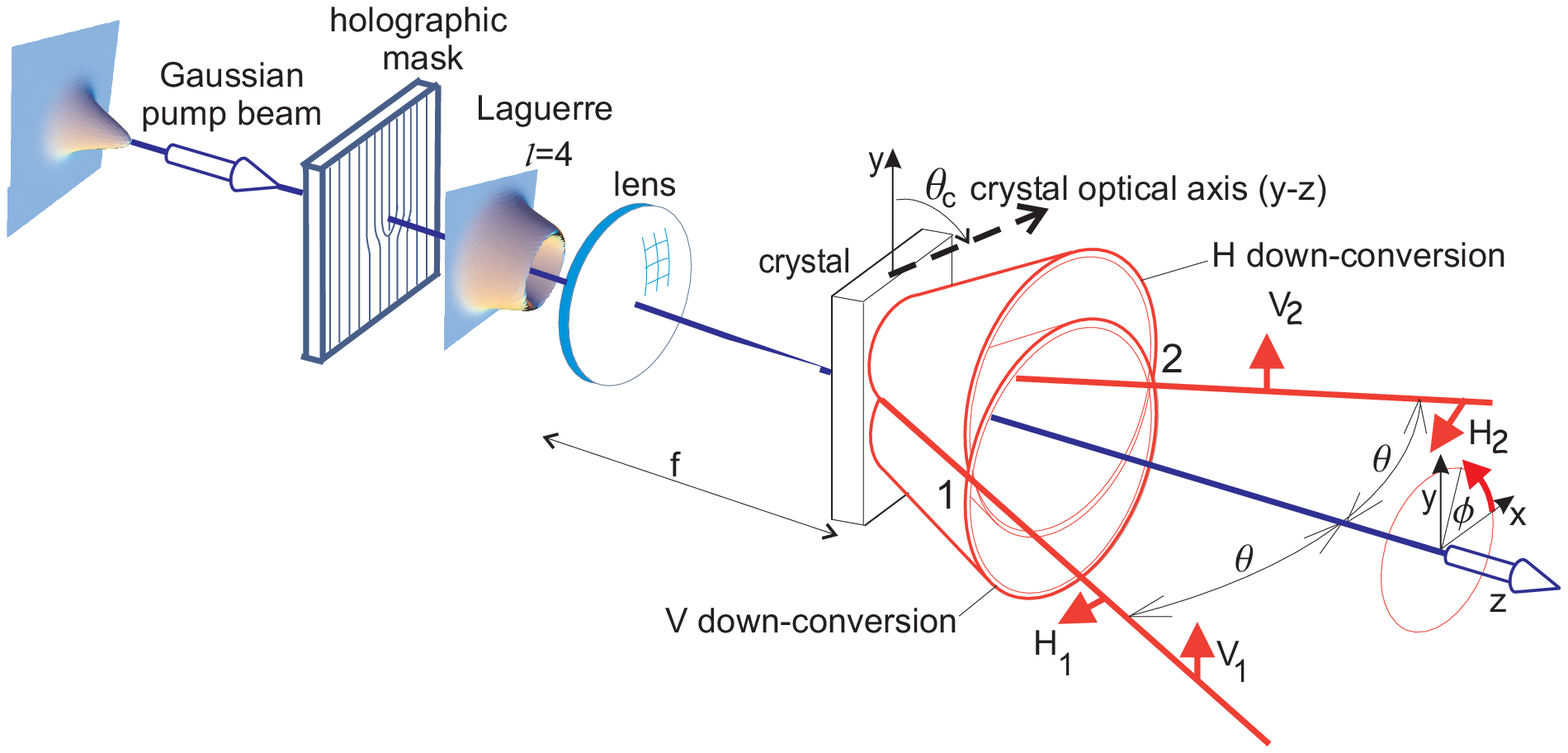}
  \end{picture}
 \end{minipage}
 \end{center}
\caption{\label{setup} (color online.) A schematic of the experimental setup. A Gaussian-profile continuous-wave argon-laser beam at 351.1nm passes through a holographic phase mask to produce a Laguerre-Gaussian beam with OAM ($l=4$). By use of an $f=50$cm focal-length convergent quartz lens, the donut-shaped beam (the measured transverse profile is shown in the lower left corner) is weakly focused into a 2mm-long beta-barium-borate (BBO) crystal cut for type-II phase matching ($\theta=49^{\circ}$, $\phi=120^{\circ}$, and the crystal is slightly tilted so that the double-ring has two separate crossings). Coincidence detections are performed to measure the coincidence images of the down-converted beams with two photon-counting modules (PCMs), both of which have a detection area of 175$\mu$m diameter and are placed after optical interference filters centered at 702nm (3nm bandwidth). One PCM is fixed at trajectory 1 while the other scans around trajectory 2 on a plane vertical to that trajectory. In all cases, the distance between the BBO crystal and the fixed (scanning) PCM is 40cm (55cm).}
\end{figure}

\begin{figure}
\vspace{0.5in}
\centerline{\scalebox{0.15}{\includegraphics{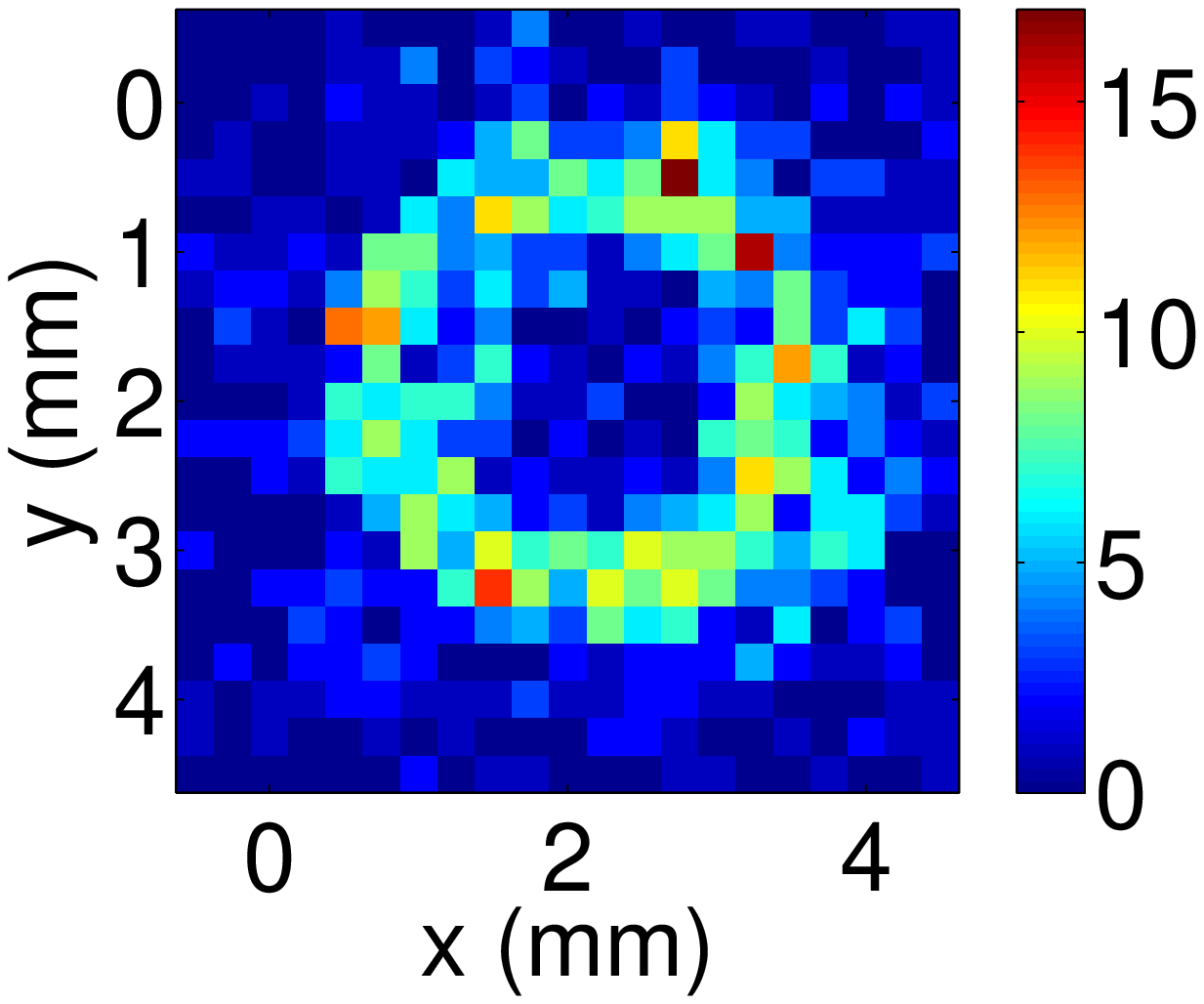}}
                           \hspace{0.0in}
            \scalebox{0.15}{\includegraphics{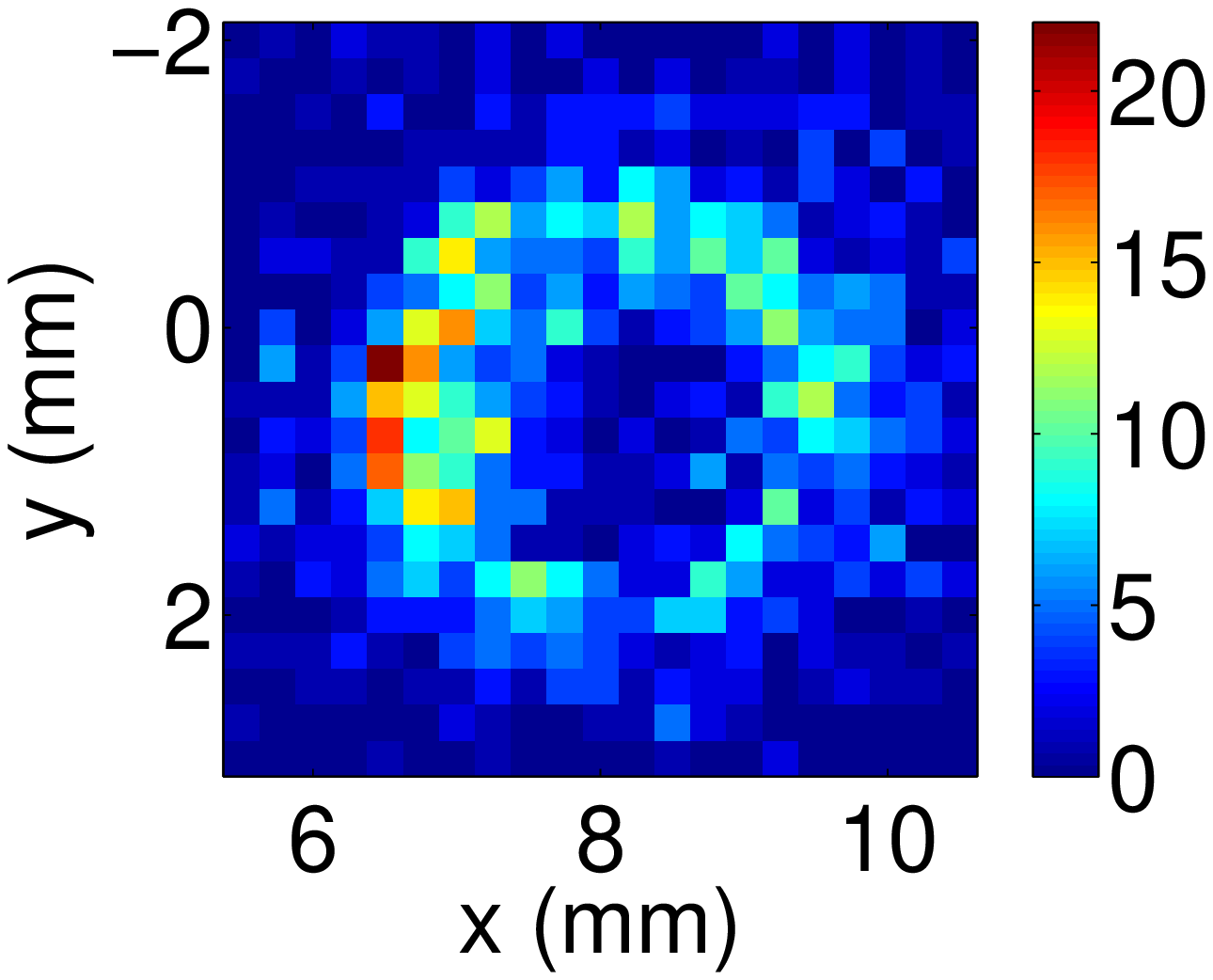}}
                           \hspace{0.0in}
            \scalebox{0.15}{\includegraphics{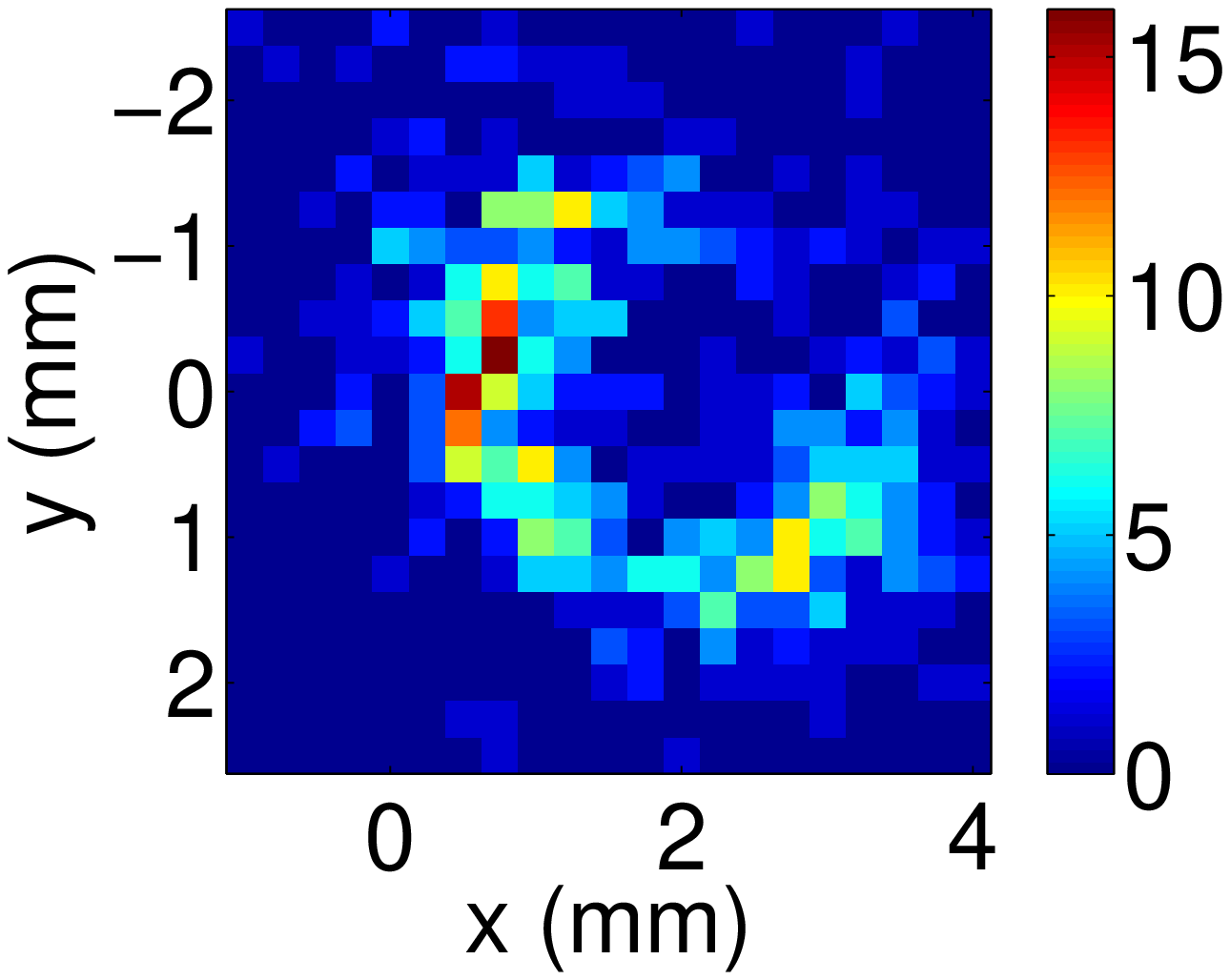}}
}
\vspace{0.1in}
\centerline{\scalebox{0.15}{\includegraphics{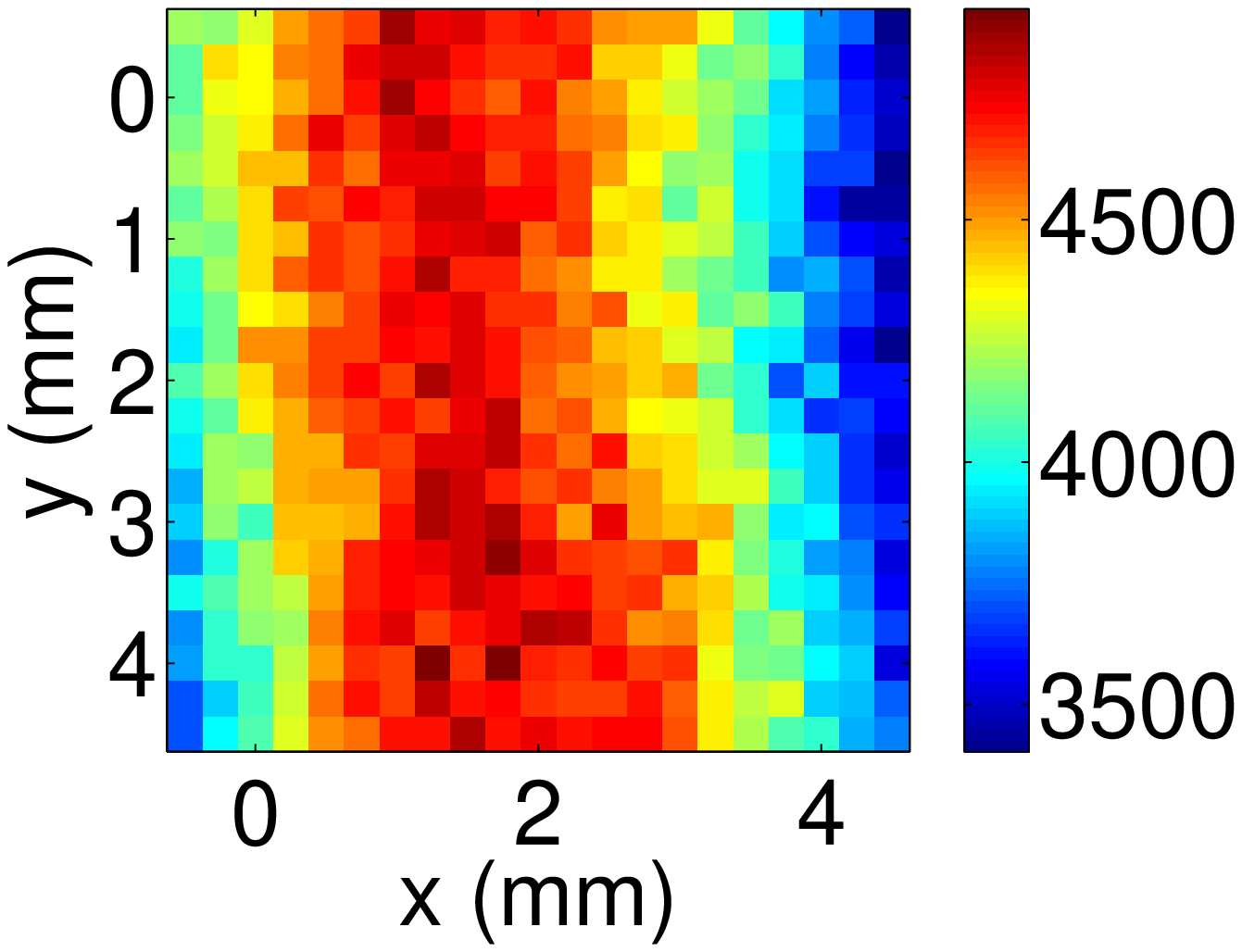}}
                           \hspace{0.0in}
            \scalebox{0.15}{\includegraphics{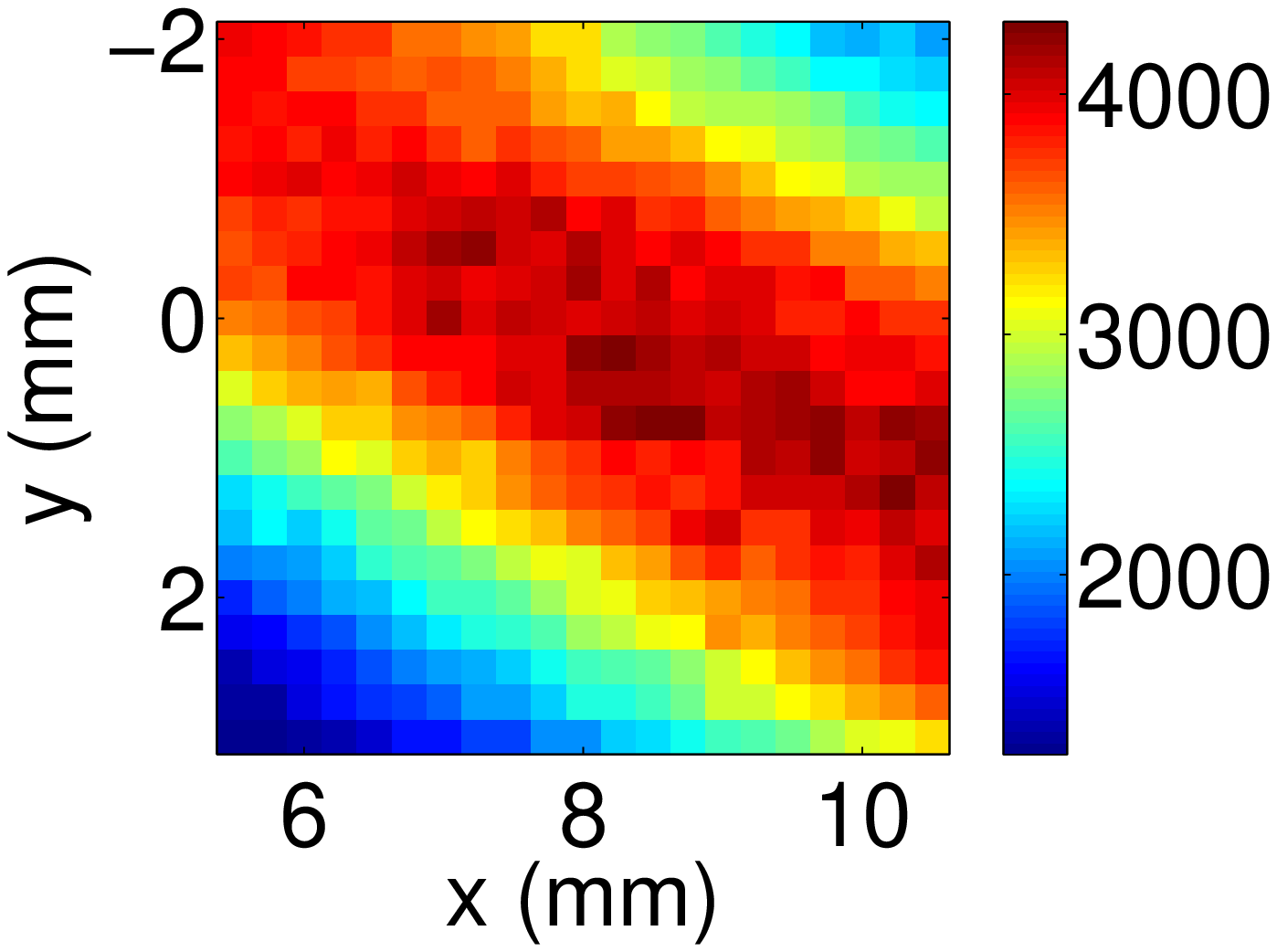}}
                           \hspace{0.0in}
            \scalebox{0.15}{\includegraphics{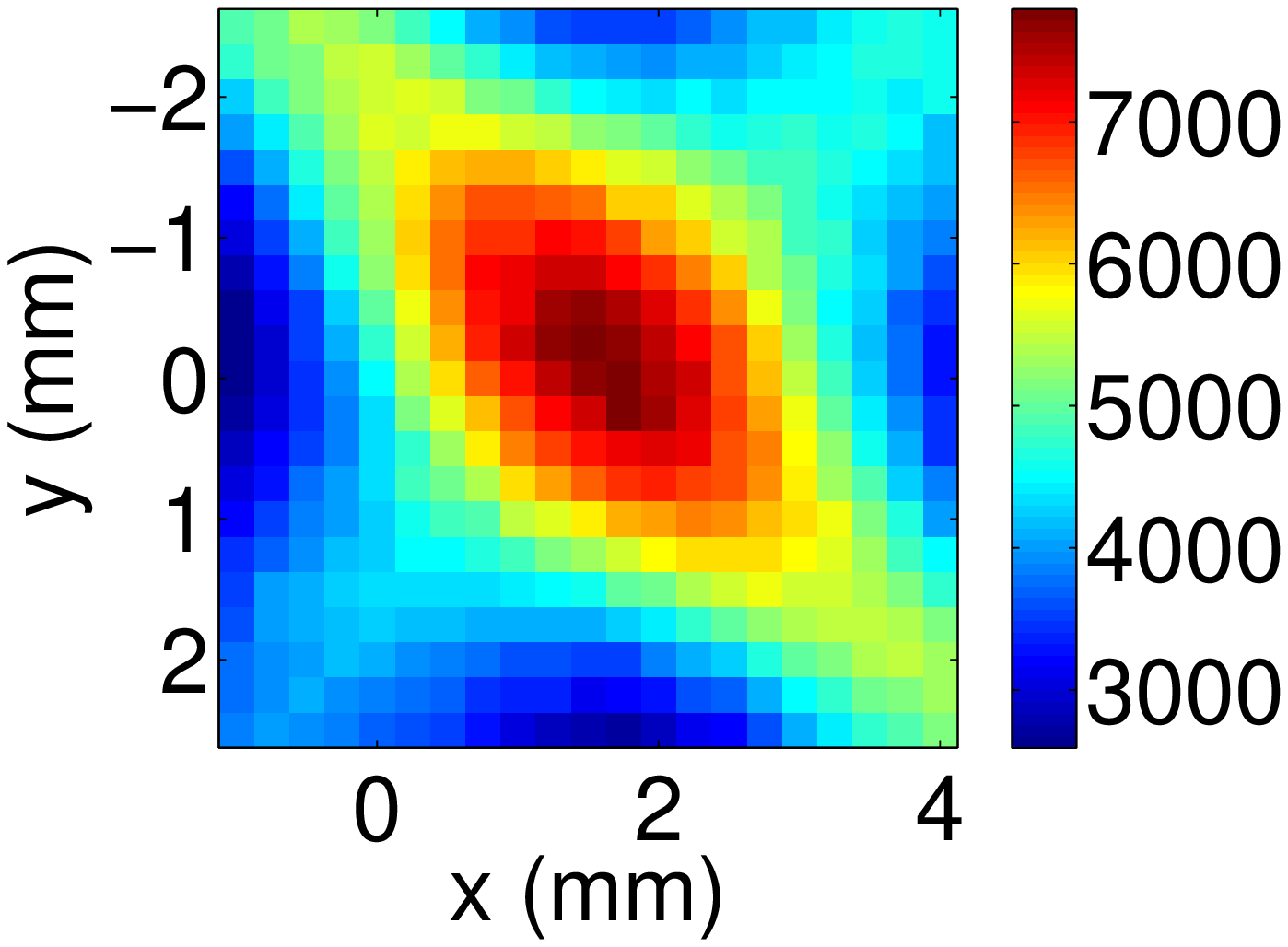}}
}
\vspace{-2.5in}
\begin{center}
\unitlength = 1mm
\begin{minipage}[c]{1mm}
 \begin{picture}(60,130)
   \hspace{-1.7in}
   \includegraphics[scale=0.4]{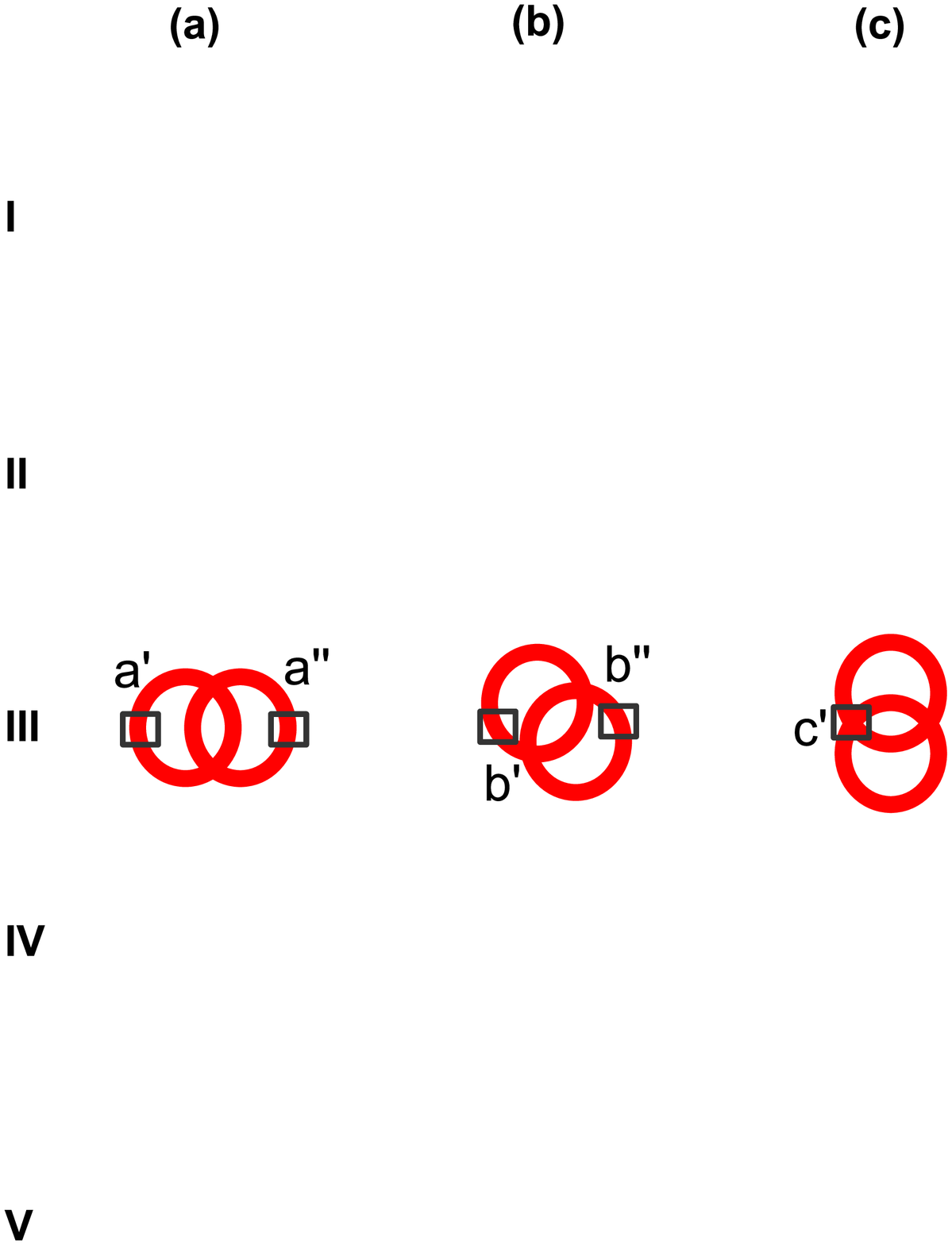}
 \end{picture}
\end{minipage}
\end{center}
\vspace{-2.0in}
\centerline{\scalebox{0.15}{\includegraphics{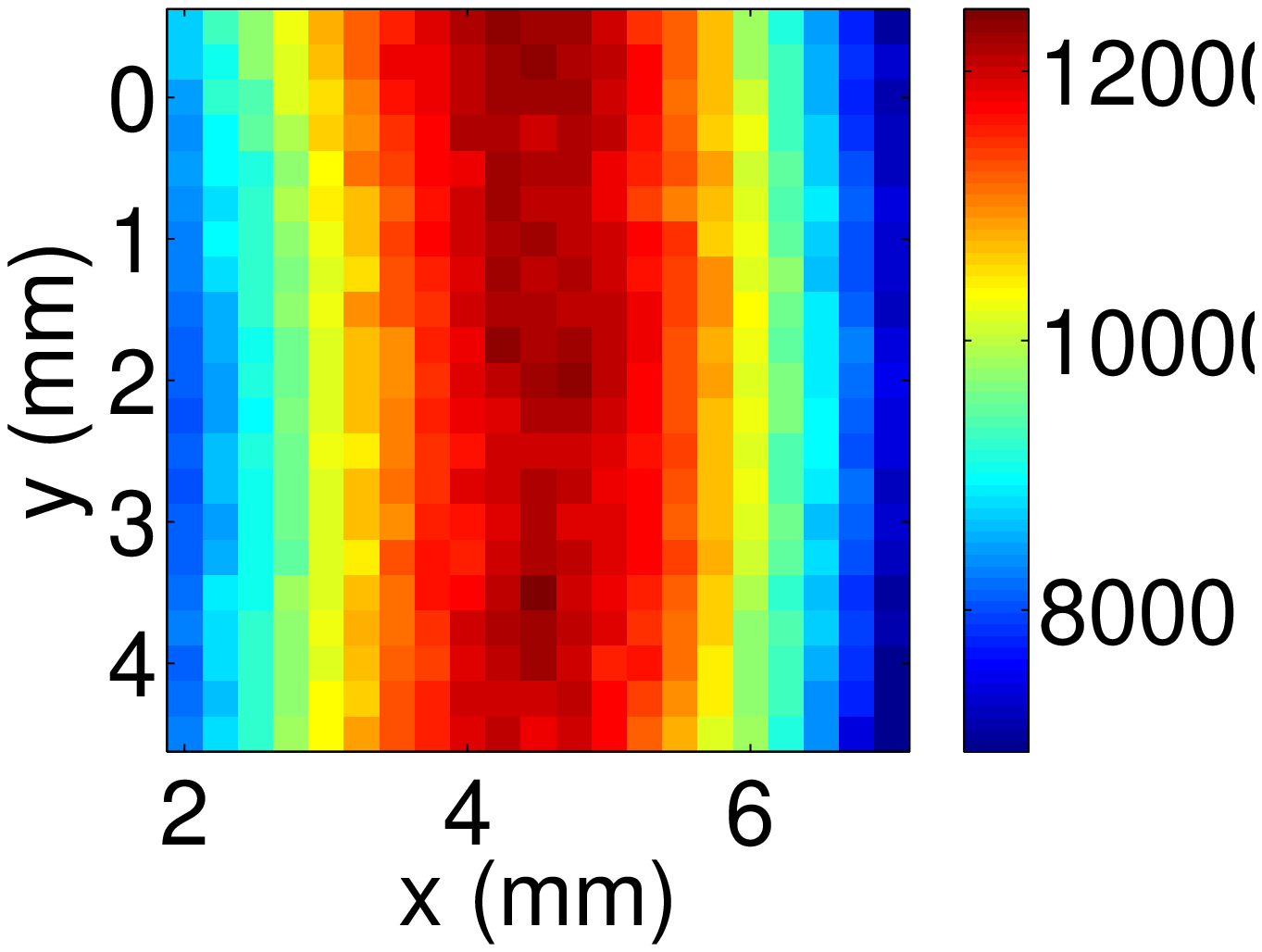}}
                           \hspace{0.0in}
            \scalebox{0.15}{\includegraphics{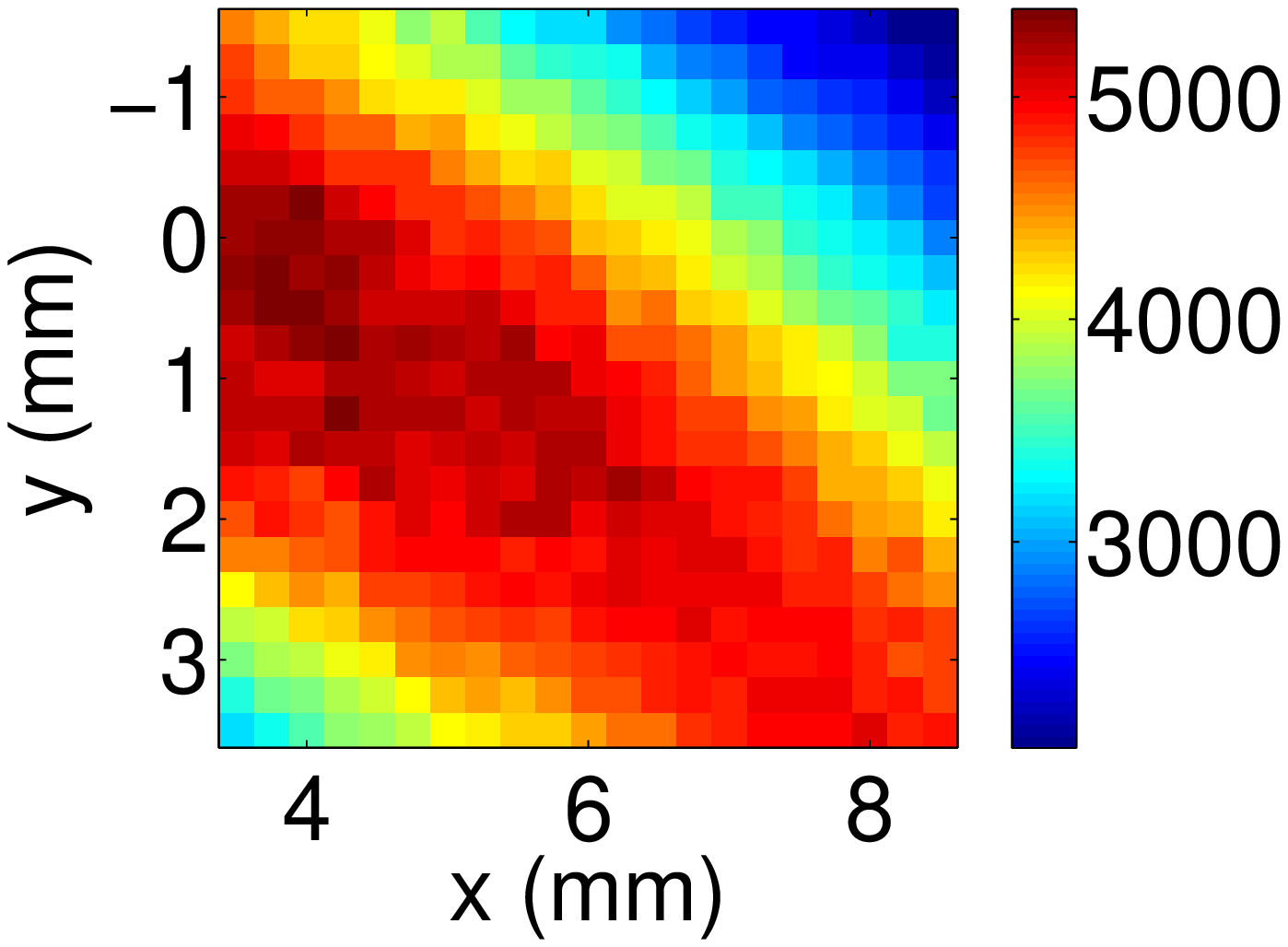}}
                           \hspace{0.0in}
            \scalebox{0.15}{\includegraphics{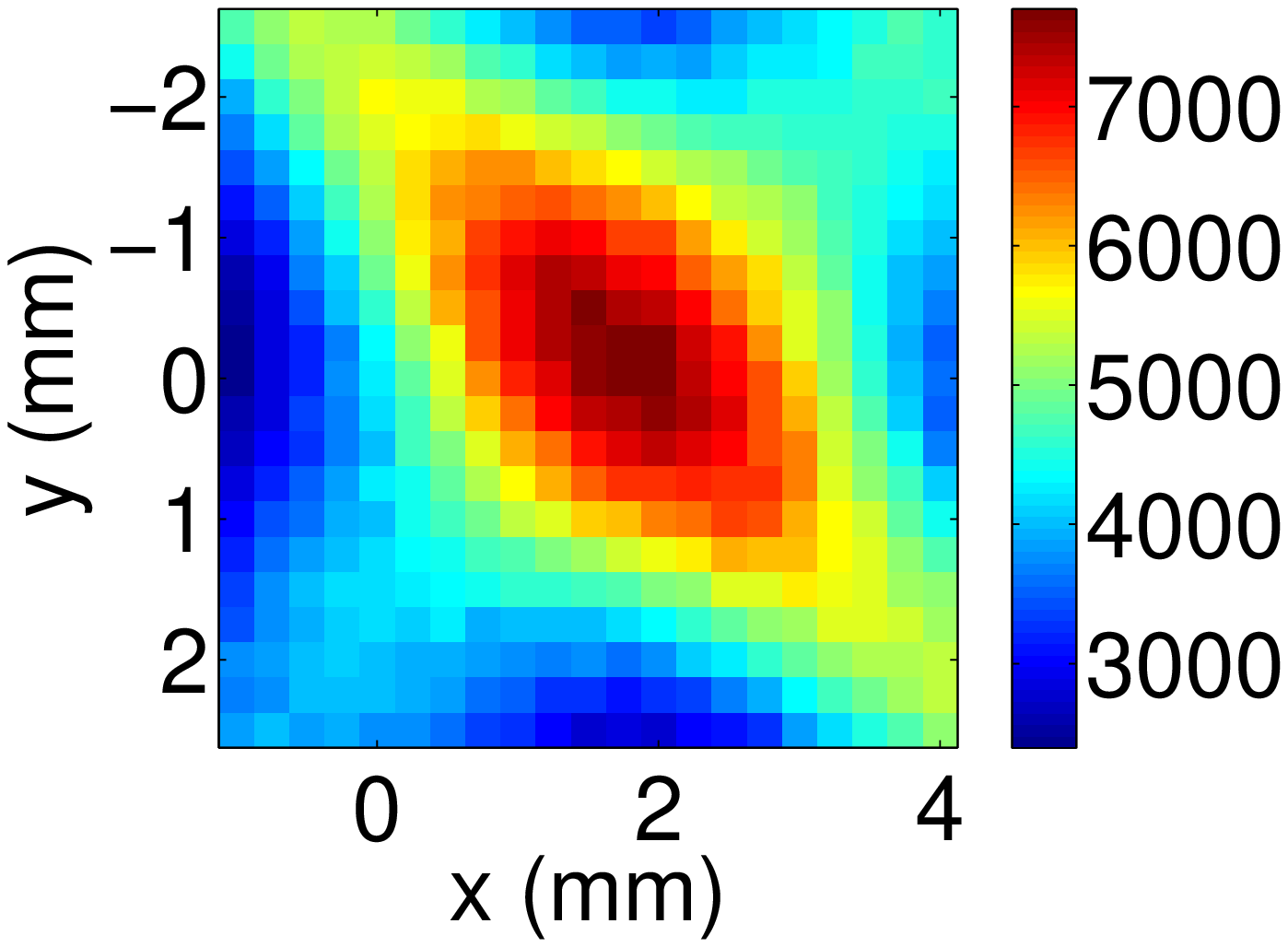}}
}
\vspace{0.1in}
\centerline{\scalebox{0.15}{\includegraphics{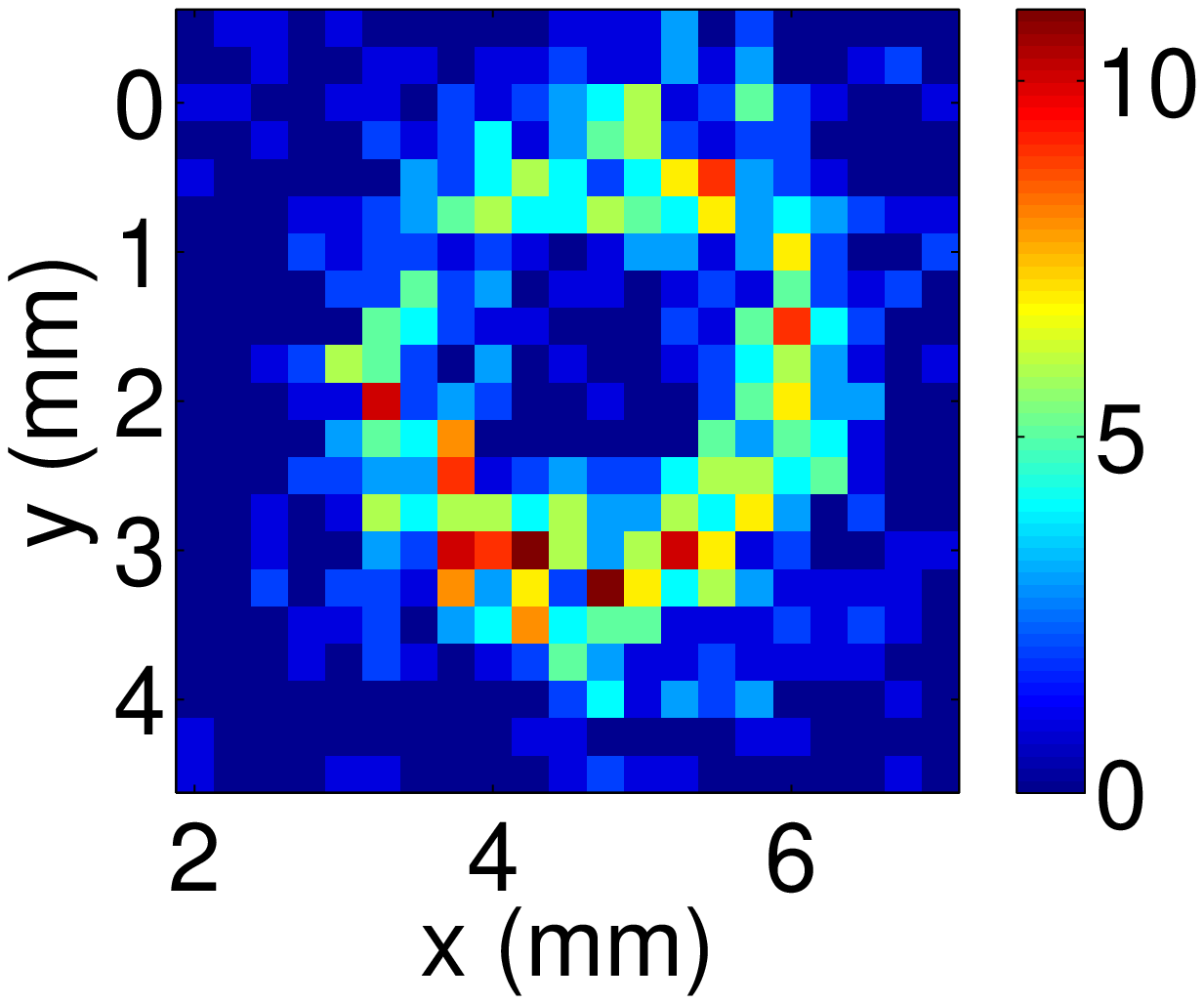}}
                           \hspace{0.0in}
            \scalebox{0.15}{\includegraphics{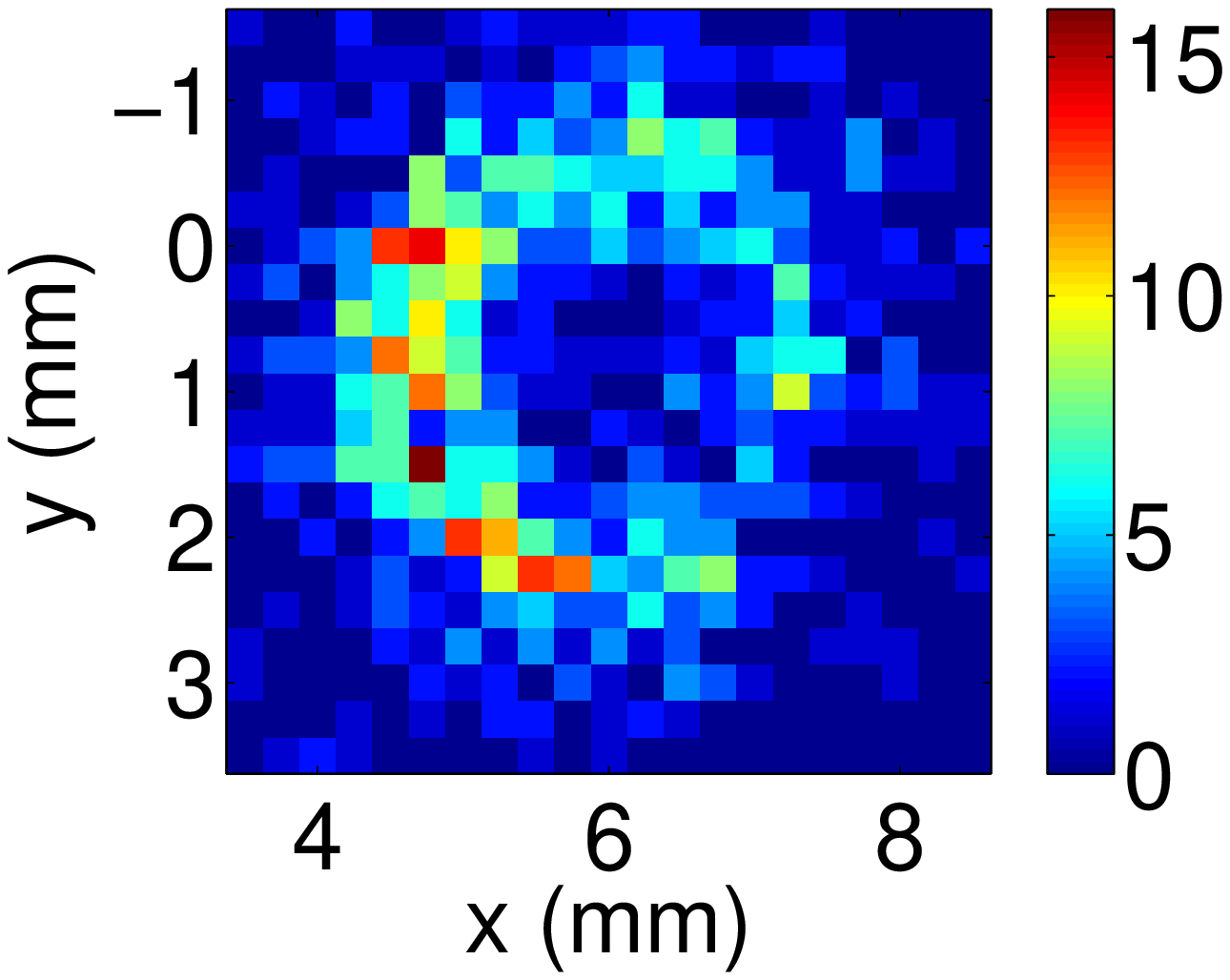}}
                           \hspace{0.0in}
            \scalebox{0.15}{\includegraphics{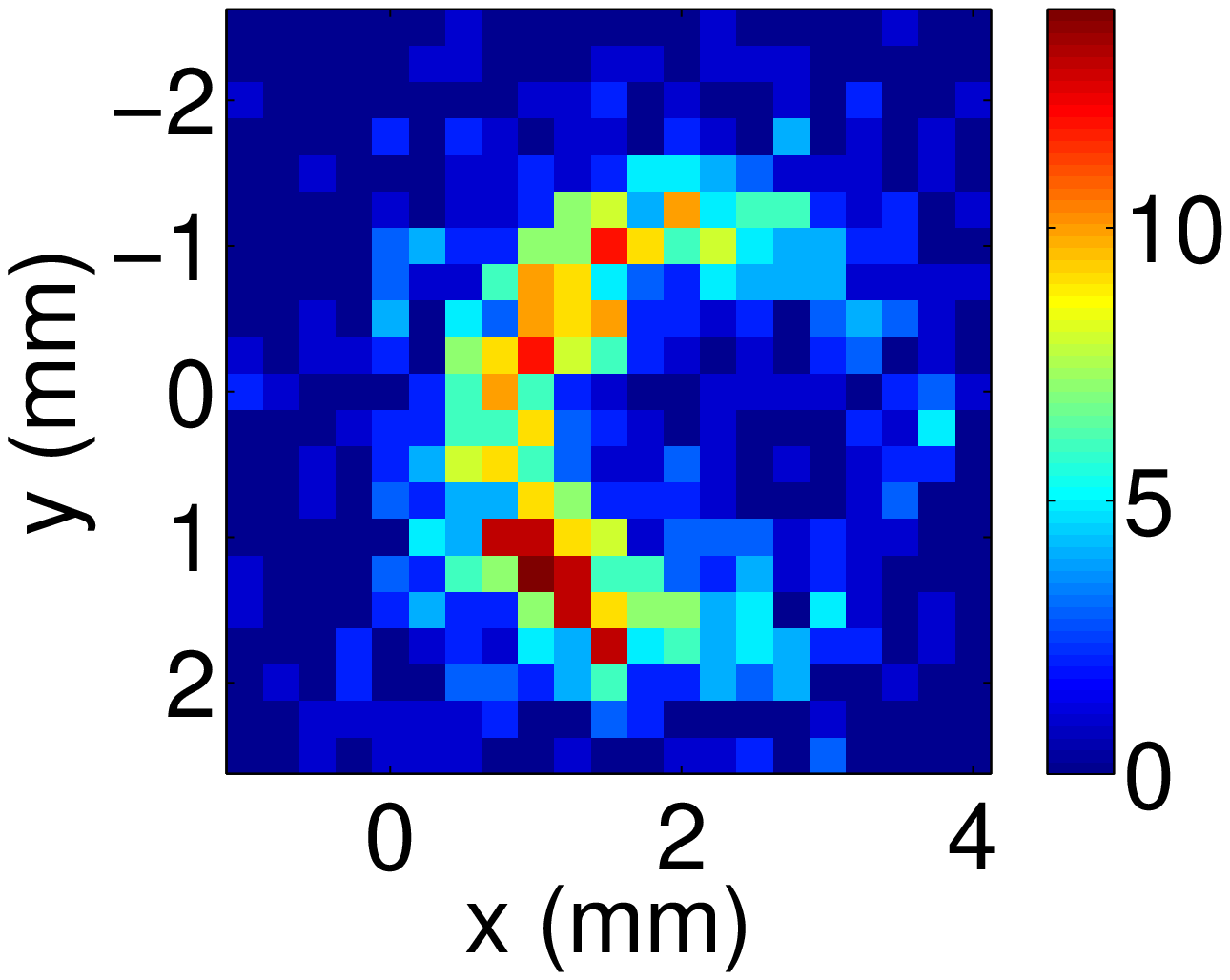}}
}
\caption{\label{data} (color online.) Coincidence images of the down-converted beams and positions of the scanning PCM. {\bf I} ({\bf V}) Coincidence images scanned on the ring of ordinary (extra-ordinary) down-converted beam over 20-30 seconds with 500mW of pump power. {\bf II} ({\bf IV}) Corresponding single counts for the ordinary (extra-ordinary) down-converted beam. {\bf III} Cartoons showing positions of the scanning PCM when coincidence detections were performed. The unbalanced single counts at a$'$ and a$''$ were due to the residual pump-focusing effect, the background from scattered pump light, and longer data-taking time at a$''$ compared to a$'$. (a) The photon pairs were collected at (a$'$, a$''$). Donut-like images are the consequence of approximate OAM conservation along the pump propagation direction ($z$-axis). (b) Data were taken after the crystal and the pump polarization had been both rotated by $45^\circ$ around $z$-axis. The coincidence images start to lose symmetry around the centers compared to those in (a), which is a signature of OAM non-conservation at (b$'$, b$''$) \cite{note2}. (c) Data were taken after the crystal was rotated by another $45^\circ$. Dramatically enhanced asymmetry shows up in the coincidence images, as a consequence of increased OAM non-conservation along $z$-axis \cite{note2}. In this case, a polarizer was used to select photons out of the ordinary (extra-ordinary) beam (i.e., no detector switching).}
\end{figure}

It was theoretically shown \cite{walborn04}, under the paraxial approximation, that OAM is conserved in the SPDC process for thin non-linear media if one assumes that the two-photon detection amplitude of the down-converted beams in the SPDC process reproduces the transverse profile of the pump beam. This is also true conversely \cite{note2}, i.e, under the paraxial approximation, the two-photon detection amplitude reproduces the transverse profile of the pump beam if OAM is conserved in the SPDC process. Because of the intrinsic connection between OAM conservation and the two-photon detection amplitude in the SPDC process, which carries information about whether OAM is conserved in the SPDC process, we performed an experiment, instead of directly measuring the OAM of each beam \cite{mair01}, to measure the moduli of the two-photon detection amplitudes \cite{note1} (coincidence images) of the down-converted beams (Fig.~\ref{setup}). Shown in Fig.~\ref{data} are our experimental results that the two-photon detection amplitude of the down-converted beams does not, in general, reproduce the transverse profile of the pump beam, which, to the best of our knowledge, can be explained only by invoking the quantum theory that conservation of angular momentum (AM) arises from rotational symmetry of the Hamiltonian governing the studied physical process, which in principle allows AM non-conservation in a phyiscal process described by a Hamiltonian with broken symmetry.

Before diving into detailed analysis, one should note that, in the type-II SPDC process, the total AM ${\bf J}$ of light is conserved {\it if and only if} the OAM ${\bf L}$ is conserved. The proof is as follows: Under the paraxial approximation, the total AM ${\bf J}$ of light can be decomposed into the sum of two separate parts: spin AM ${\bf S}$ and orbital angular momentum (OAM)  ${\bf L}$ \cite{barnett02}, i.e., ${\bf J}={\bf S}+{\bf L}$. In our experiment, the pump and the down-converted photons are linearly polarized. Then the spin AM ${\bf S}$ is conserved in the SPDC process since photons with linear polarization carry spin AM with defined value, which is zero. This conservation can be confirmed by noting that $[{\bf S}, H]=0$ in the Hilbert sub-space spanned by the initial state vector in which the down-converted modes are empty and the final state vector in which each down-converted mode is occupied by one photon with linear polarization. Here ${\bf S}$ is the total spin operator of the light beams and $H$ is the Hamiltonian ruling the non-linear process with a linearly polarized pump-beam. Therefore, ${\bf J}$ is conserved if ${\bf L}$ is conserved, and vice versa.

\begin{figure}
\includegraphics [scale=0.3]{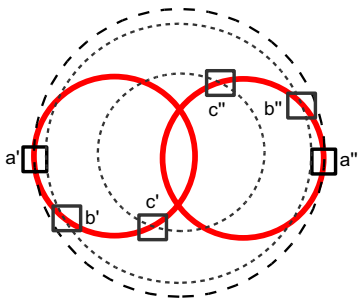}%
\caption{\label{sym} (color online.) Local symmetric properties of down conversion in type-II SPDC. The down-conversion (red solid curves) at position a$'$ and a$''$ can be approximated by a single ring (coarse dashed curve) that has perfect azimuthal symmetry. At position b$'$ and b$''$ the single-ring approximation becomes poor, and poorer in c$'$ and c$''$ (fine dashed curves). Detailed numerical simulation for the symmetric properties of the corresponding Hamiltonian, which agrees with the analysis shown here, will be presented elsewhere as part of a theoretical work.}
\end{figure}

To explain the experimental results, we analyze the symmetric properties of the Hamiltonian of the type-II SPDC process as follows. The spatial pattern of down conversion in the type-II SPDC process is a double-ring structure (Fig.~\ref{sym}) that does not possess rotational symmetry around the pump propagation direction (azimuthal symmetry around $z$-axis). In other words, the average rate of down-conversion, $R \propto \langle \psi(t)|a^{\dagger}a|\psi(t)\rangle$, is a function dependent of the azimuthal angle. This implies that the Hamiltonian $H_{II}$ governing the type-II process, unlike in the type-I case, will not remain constant under the operation of space rotation around $z$-axis. Therefore, the commutation relation $[J_z, H_{II}]\ne0$ ( $J_z$ is the $z$-component of the AM of light), which means that $J_z$ is not a good quantum number and not conserved by definition. The key point is that the degree of azimuthal symmetry breaking around $z$-axis and, consequently, AM (OAM) non-conservation along $z$-axis in the type-II process depend on where the photon pairs are created on the double-ring. For example, if photon detectors are set to collect photon pairs at position a$'$ and a$''$, as depicted in Fig.~\ref{sym}, where the double-ring can be approximated by a single symmetric ring, the azimuthal symmetry breaking is minimal and may be negligible. In this case, AM (OAM) non-conservation along $z$-axis could be small compared to experimental resolution and approximate AM (OAM) conservation may exist. On the other hand, if photon pairs are detected at position b$'$ and b$''$ (Fig.~\ref{sym}), where the double-ring can hardly be approximated by a single symmetric ring, the azimuthal symmetry breaking may not be negligible and AM (OAM) non-conservation may start to show up in experimental observation.

With the above analysis, one can explain the experimental observation very well. As shown in Fig.~\ref{data}(a), we first measured the coincidence images at positions (a$'$, a$''$), where the non-linear process possesses approximate rotational symmetry and approximate OAM conservation along $z$-axis holds, which is consistent with published results \cite{walborn04}. Consequently, the two-photon detection amplitude is similar to the transverse profile of the pump beam \cite{note2}, agreeing with the observation. Then, the photon detectors were re-aligned (actually, the non-linear medium and the polarization of the pump were rotated by some angle around the pump axis for practical convenience) to collect photon pairs at positions, as an example, (b$'$, b$''$) in Fig.~\ref{data}(b). In this case, the coincidence images have observable difference from the transverse profile of the pump beam, as a consequence of non-negligible OAM non-conservation \cite{note2}. Continuing this way, we obtained a series of coincidence images, each pair of which was measured at different positions [Fig.~\ref{data}(a)-\ref{data}(c)]. The first pair of images can be considered as reference; the pairs measured at subsequent positions are increasingly asymmetric around the centers compared to the reference images, which is correlated to the degree of rotational symmetry breaking at these posistions where data were taken according to the above qualitative analysis! This suggests OAM non-conservation along $z$-axis in the type-II SPDC process caused by rotational symmetry breaking of the Hamiltonian.

In these successive measurements [Fig.~\ref{data}(a)-~\ref{data}(c)], all other experimental parameters remain unchanged and their errors, if any, were common mode and cancel out (the reference images carry the same experimental errors as other images, which is a major advantage of our experimental design) except the parameter associated with the photon-pair collection position, which determines two physical quantities: photon-pair-emission polar angle and the degree of rotational symmetry breaking. To eliminate the possibility that our observation was the consequence of OAM non-conservation caused by the non-collinear geometry \cite{terriza03,vaziri03,torres03}, we note that the photon-pair-emission polar angles in all the measurements were less than $10^{\circ}$, which is within the limit of $18^{\circ}$ given in \cite{terriza03} beyond which OAM non-conservation was said to be experimentally visible. Furthermore, in the case where coincidence images were broken to the most extent in these measurements, as shown in Fig.~\ref{data}(c), the polar angle was less than $4^{\circ}$! One should recall that, at a polar angle of $4^{\circ}$, OAM conservation was observed in the type-I SPDC process\cite{mair01,vaziri02}. Hence our observation can only be connected to the rotational symmetry breaking.

Regarding the interpretation of data, one must also consider the effect of pump focusing, which can affect the coincidence image of the down-converted beams \cite{lee05,torres05} and invalidate the local paraxial approximation \cite{walborn04,huguenin06}. In our experiment, however, we recall that the pump was weakly focused by a 50cm focal-length lens that produced a beam with Rayleigh length of about 10cm, which is much greater than 2mm, the length of the non-linear medium. In addition, we note that, if the pump-focusing effect is a dominant factor that causes asymmetry in the coincidence images, for example, at the double-ring crossing [position c$'$ in Fig.~\ref{data}(c)], then a dramatic asymmetry should also be present in the coincidence images scanned at position a$''$ on the extra-ordinary beam. The donut-shaped coincidence images at positions a$'$ and a$''$ indicate that the pump-focusing effect alone, which can never be completely eliminated in practice though, was not strong enough to cause the observed asymmetry in other coincidence images in Fig.~\ref{data}. So, the experimental errors introduced by pump focusing do not invalidate our arguments.

When arguing that AM is not conserved in the type-II SPDC process due to broken spatial symmetry of the Hamiltonian, one is faced with a straightforward question: Where did the AM go? Intuitively, one might think of non-zero AM exchange between the interacting light waves and the non-linear media. Then the total AM of the ``isolated'' system including light beams and media may still be conserved. This, given that AM conservation arises from symmetry of the Hamiltonian, is equivalent to state that the Hamiltonian with broken symmetry, which is traditionally used to describe the SPDC process to the first order approximation \cite{louisell61,klyshko69,burham70,ou89,arnaut00}, is not complete and that neglected terms need to be explicitly written for a complete Hamiltonian with perfect symmetry. But the challenge is to show that dropped terms can really compensate the symmetry breaking of the existing terms. In addition, the microscopic mechanism through which non-zero AM exchange, if any, between light and media takes place in the studied process also needs to be further investigated both theoretically and experimentally.


While welcoming all efforts from the literature to fully explore this field, we note that our experiment, together with the explanation, may open a new way to practically preserve OAM conservation in the SPDC process, as shown in Eq. (12) in \cite{arnaut00}. According to this equation, the azimuthal symmetry of the Hamiltonian is directly related to the product of two terms: the interaction amplitude $A_{{\bf k},s;{\bf k}',s'}$ and the spatial phase-matching $\tilde{\psi}_{lp}(\Delta {\bf k})$. While $A_{{\bf k},s;{\bf k}',s'}$ depends both on the non-linear susceptibility $\chi^{(2)}$ and the linear susceptibility $\chi^{(1)}$, the latter defining the unit polarization vectors in the medium through the linear refractive indices, the $\tilde{\psi}_{lp}(\Delta {\bf k})$ defines the phase-matching directions whenever $A_{{\bf k},s;{\bf k}',s'}$ is non-zero. Some experimental tools can be used to induce controlled medium modifications. Pressure, electric and magnetic fields are natural choices to exert different modifications on the medium, whereas techniques such as quasi-phase-matching \cite{torres04} can be used to change the phase-matching condition. Through these parameters, practically feasible tools could be invented to control the symmetric properties of the Hamiltonian and engineer the OAM conservation for high-quality OAM entanglement.

In conclusion, we experimentally discovered that the two-photon detection amplitude of the down-converted beams does not reproduce the transverse profile of the pump beam, which shows OAM non-conservation, in the type-II SPDC process. This can be very well explained by the broken rotational symmetry of the Hamiltonian, which is fundamentally connected to AM (OAM) non-conservation in the studied process. However, to thoroughly understand the physics involved in our experiment, more theoretical and experimental investigation is on demand. 

This work was supported by the U.S. Army Research Office Multidisciplinary University Research Initiative Grant on Quantum Imaging. We thank A. Vaziri and A. Zeilinger for the holographic mask in use.


\begin{thebibliography}{}
\bibitem{note1} As a matter of fact, what was measured was the squared modulus of the two-photon detection amplitude (the photon-pair generation rate) multiplied by the measurement time, i.e., the coincidence-count distribution on the transverse plane.
\end{thebibliography}
\end{document}